\documentclass[structabstract]{raa}

\usepackage{soul}
\usepackage{color}
\usepackage{booktabs}
\usepackage{graphicx,times}
\usepackage{natbib,amssymb}
\usepackage{epstopdf,epsfig}
\usepackage{float}
\usepackage{amsmath}
\usepackage{ulem}
\usepackage{amsmath}
\usepackage{hyperref}
\usepackage{mathtools}

\providecommand{\bm}[1]{\mbox{\boldmath$#1$\unboldmath}}

\begin{document}
 \title{The Formation of Electron-capture Supernovae: A Review}

   \volnopage{ {\bf 2026} Vol.\ {\bf X} No. {\bf XX}, 000--000}
   \setcounter{page}{1}
   \author{Bo Wang\inst{1,2,3},
	       Dongdong Liu\inst{1,2,3},
	       Yunlang Guo\inst{4,5},
          \and
           Zhanwen Han\inst{1,2,3}
          }
   \institute{Yunnan Observatories, Chinese Academy of Sciences, Kunming 650216, China;
          {\it wangbo@ynao.ac.cn; liudongdong@ynao.ac.cn; yunlang@nju.edu.cn; zhanwenhan@ynao.ac.cn}\\
          \and
          International Center of Supernovae, Yunnan Key Laboratory, Kunming 650216, China\\
          \and
          University of Chinese Academy of Sciences, Beijing 100049, China\\
          \and
         School of Astronomy and Space Science, Nanjing University, Nanjing 210023, China\\
          \and
          Key Laboratory of Modern Astronomy and Astrophysics, Nanjing University, Ministry of Education, Nanjing 210023, China\\
              }

\date{Received ; accepted}

\abstract {It is generally believed that
the electron-capture reactions happen
when the oxygen-neon (ONe) cores grow in masses close to the Chandrasekhar limit,
leading to the formation of neutron stars (NSs) via electron-capture supernovae (EC-SNe). 
EC-SNe are predicted to be the most likely short-lived and faint optical transients, and  
a small ejecta mass is expected during the collapse. This kind of SNe provide a distinct channel for producing isolated NSs and NS systems, 
especially for the formation of X-ray binaries and double NSs.
Although EC-SNe were proposed $\sim$45\,yr ago, there are still some uncertainties for the origin of EC-SNe and their productions.
In this article, we review recent studies on the two classic progenitor channels of EC-SNe, i.e., the single star channel and the binary channel. 
In the single star channel, EC-SNe can happen in super asymptotic giant branch stars or He stars,
whereas in the binary channel EC-SNe can occur in He stars in binaries 
(involving He star+MS systems and NS+He star systems) or accretion-induced collapse in white dwarf binaries 
(involving the single-degenerate scenario and the double-degenerate scenario).
Recent progress on these two progenitor channels is discussed, 
including the initial parameter range for EC-SNe, the evolutionary paths to EC-SNe, related objects and some observational constraints, etc. 
We also make some discussions on the possible candidates for EC-SNe in this article,
and the impacts of EC-SNe on some research fields, e.g., the properties of NSs, 
double NS population and chemical products, etc.
It is noting that EC-SNe show some similar properties with ultra-stripped SNe, e.g., 
low ejecta masses and small kicks. Accordingly, we also discuss
the difference between these two types of SNe in this article.
Research on EC-SNe is at a pivotal stage, with key theoretical uncertainties and observational challenges requiring integrated modeling and multi-wavelength observations for robust identification.
 }

\titlerunning{Electron-capture Supernovae}

\authorrunning{B. Wang et al.}

\maketitle

\section{Introduction}
Supernovae (SNe) are violent stellar explosions in the Universe, which occur when a star dies (see \citealt{1960ApJ...132..565H}). 
They play an important role in modern astrophysics, as follows:
(1) They have been used as the cosmological distance indicators, revealing the discovery of the accelerating expansion of the Universe (see \citealt{Riess1998AJ,Perlmutter1999ApJ}). 
(2) They play a prominent role in understanding of galactic chemical evolution due to the main contribution of 
heavy elements to their host galaxies, especially for both intermediate mass elements and iron-group elements  
(see \citealt{Greggio1983,Matteucci1986}).
(3) They are kinetic-energy sources in galaxy evolution and important cosmic-ray accelerators (see \citealt{Helder2009Sci,Powell2011MNRAS,LiY2018ApJ}). 
(4) Their observations can be used to test stellar evolution theory and to constrain star formation history of the Universe
(see \citealt{Cappellaro1999,Mannucci2006MNRAS,Maoz2014}). 
(5) During the SN explosion, they can be used to test some fundamental physics, such as the detections of neutrinos, gravitational waves and shock breakout, etc (e.g., \citealt{Li2024Natur,Zhang2024ApJ}).   

SNe represent the final stage for the certain types of stellar evolution, 
including explosions of massive stars and white dwarf (WD) systems. 
They are mainly categorized into two physically distinct classes, i.e., thermonuclear explosion SNe (type Ia SNe) 
and core-collapse SNe (type Ib, Ic, IIP, IIL, IIn, IIb SNe)  (see, e.g., \citealt{Filippenko1997,Parrent2014}). 
Type I SNe are distinguished by the absence of H lines in their spectra, whereas type II SNe  show obvious H lines. 
\citet{Ma2025A&A...698A.305M} recently established a complete sample of 211 SNe within 40 Mpc (with a mean distance of $~$26 Mpc), deriving fractions of type Ia SNe  (30.4\%), type Ib/c SNe (16.3\%) and type II SNe (53.3\%). 
Thermonuclear explosion SNe will blow up the whole object when the SN explodes, 
which originate in WD binary systems (see, e.g., \citealt{Wang2012NewAR..56..122W, Maoz2014,Wang2018RAA,Liu2023RAA....23h2001L,Ruiter2025A&ARv..33....1R}).
Core-collapse SNe originate from the evolution of massive stars, which will be left with a neutron star (NS) or a black hole (BH) after the SN explosion  (see, e.g., \citealt{Smartt2009ARA&A, Burrows2024ApJ...964L..16B}).
According to the difference of explosion mechanisms, core-collapse SNe can be divided into 
electron-capture (EC) SNe and iron core-collapse (Fe CC) SNe. 
\citet{Hiramatsu2021NatAs...5..903H} estimated that about $0.6\%-8.5\%$ of all core-collapse SNe belong to EC-SNe. 

EC-SNe are thought to arise from electron-capture reactions on $^{24}\rm Mg$ and $^{20}\rm Ne$ in the degenerate oxygen-neon (ONe) cores with masses close to the Chandrasekhar limit ($M_{\rm Ch}$), in which the ONe cores eventually collapse into NSs (see \citealt{Miyaji+1980,Nomoto1982Natur.299..803N,Nomoto1984ApJ,Nomoto1987ApJ,Doherty2015MNRAS.446.2599D,Nomoto2017hsn..book..483N,Zha2019ApJ...886...22Z,Zha2022MNRAS,Tauris2023pbse.book.....T}).
In the classic picture,
the ONe cores for producing EC-SNe have masses  $\sim1.37-1.43\,M_\odot$  with a relatively high ignition density ($\rm log_{10}(\rho_{\rm c}/g\,cm^{-3})\gtrsim10$; see, e.g., \citealt{Nomoto1984ApJ,Takahashi2013ApJ...771...28T,Tauris2015MNRAS.451.2123T,Guo2024MNRAS.530.4461G}).
For the ONe core masses $>1.43\,M_\odot$, the stars will collapse into NSs via Fe CC-SNe,
whereas stars with ONe core masses $<1.37\,M_\odot$ will form ONe WDs. 

For the ONe cores with masses $\sim1.37-1.43\,M_\odot$, 
the oxygen deflagration starts in their center
owing to the heating of electron-capture on $^{20}\rm Ne$.
When the central temperature of the ONe core rises rapidly,
nuclear statistical equilibrium will be arrived
once the temperature exceeds $5\times10^9$\,K.
In the nuclear statistical equilibrium region, 
the electron-capture reactions accelerate the contraction of the ONe core, leading to the formation of NSs  once
the electron-capture rate exceeds the nuclear burning rate.
It is still a theoretical issue whether NSs can be produced after the oxygen deflagration happens (see \citealt{LeungShing-Chi2019PASA...36....6L}). By simulating the oxygen 
deflagration via multidimensional hydrodynamics, \citet{Jones2016} suggested that the ONe cores will not collapse into NSs when the oxygen deflagration is triggered with a low ignition density ($\rm log_{10}(\rho_{\rm c}/g\,cm^{-3})\sim9.8$), 
leaving bound ONeFe WD remnants with $\sim 0.1-1\,M_\odot$ ejecta,
called the thermonuclear EC-SNe
(see also \citealt{Isern1991ApJ...372L..83I,Jones2019A&A,Kirsebom2019PhRvL.123z2701K,Tauris2019ApJ...886L..20T}).

It has been suggested that EC-SNe are relatively faint optical transients, producing 
low explosion energies ($\rm \sim 10^{50}\,erg$), 
low masses of $^{56}\rm Ni$ ($\sim 0.002-0.015\,M_\odot$) and
small amounts of ejecta mass ($\sim 0.01-0.2\,M_\odot$) during the collapse
(see, e.g., \citealt{Timmes1996ApJ...457..834T,Kitaura2006A&A...450..345K, Wanajo2009ApJ...695..208W, Fryer2012ApJ...749...91F}).
In addition, EC-SNe are expected to produce low-velocity NS kicks
($\rm \lesssim 100\,\rm km\,s^{-1}$; see \citealt{Podsiadlowski2004ApJ...612.1044P,Gessner2018ApJ...865...61G,Stockinger2020MNRAS.496.2039S}). 
It is worth noting that the kick velocity is probably lower than $\sim50\,\rm km\,s^{-1}$, 
corresponding to the explosion energy $\sim 10^{50}$\,erg 
(e.g., \citealt{Kitaura2006A&A...450..345K,Dessart2006ApJ...644.1063D,Janka2017ApJ...837...84J}).


EC-SNe may contribute to the formation of isolated NSs and NS systems, 
especially for the production of X-ray binaries and double NSs (see \citealt{Dewi2002MNRAS.331.1027D,Tauris2015MNRAS.451.2123T, Shao2016ApJ...816...45S,Beniamini2016MNRAS.456.4089B,Kruckow2018MNRAS.481.1908K}). 
They can synthesize heavy elements through the $r$-process (e.g., \citealt{Ning2007ApJ...667L.159N}).
Meanwhile, they can contribute to lower mass peak ($\sim 1.25\,M_\odot$) in the bimodal distribution of NS masses (e.g., \citealt{Schwab2010ApJ...719..722S}).
In observations, some low luminosity SNe or events are expected to be produced through EC-SNe (see, e.g., \citealt{Takahashi2013ApJ...771...28T, Nomoto2017hsn..book..483N}).

\begin{figure}
\begin{center}
\epsfig{file=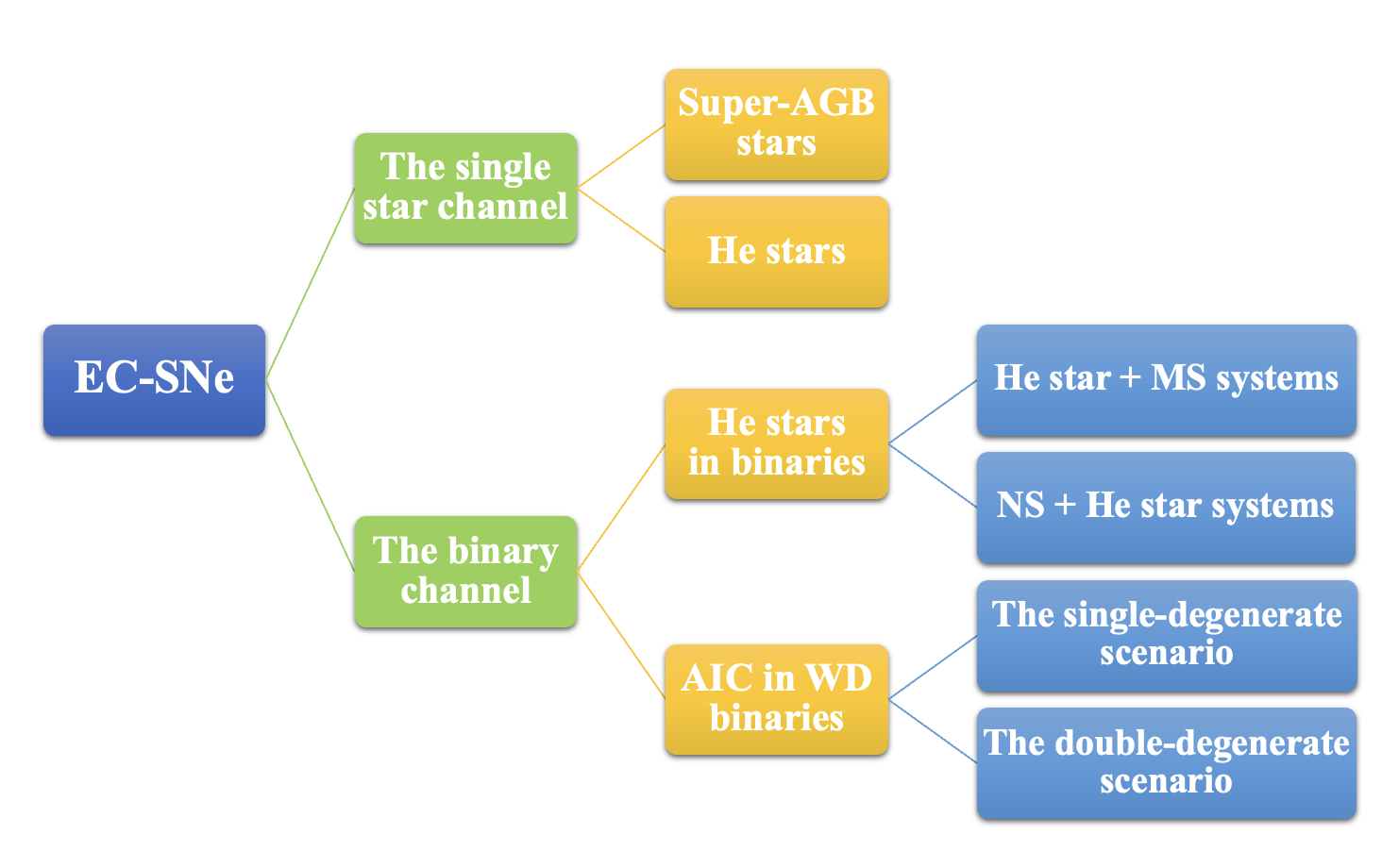,angle=0,width=12.5cm}
\caption{Evolutionary channel flowchart for the formation of EC-SNe.}
  \end{center}
  \label{fig:channel}
\end{figure}

In this review, we mainly focus on the formation of EC-SNe.
There are two primary formation channels for producing EC-SNe, i.e., 
the single star channel and the binary channel.
Fig. 1 shows a schematic diagram that summarizes all discussed pathways
for the formation of EC-SNe in this article.
In Sect. 2, we review the single star channel for the formation of EC-SNe, including super Asymptotic Giant Branch (AGB) stars and He stars.
The binary channel for EC-SNe is summarized in Sect. 3, 
including both He star in binaries (He star+MS systems and NS+He star systems)
and accretion-induced collapse (AIC) in WD binaries (the single-degenerate scenario and
the double-degenerate scenario). 
We discuss the possible candidates for EC-SNe in Sect. 4, 
the impacts of EC-SNe on some research fields 
in Sect. 5 and the difference between EC-SNe and ultra-stripped SNe in Sect. 6.
A summary and perspective are provided in Sect. 7.

\section{The single star channel}
\subsection{Super-AGB stars}

In the classic single star channel, the ONe core in a super-AGB star  will collapse into a NS 
through electron-capture reactions when it grows in mass close to $M_{\rm Ch}$ (see Fig. 2).
It has been suggested that super-AGB stars
for producing EC-SNe have initial main-sequence (MS) masses of $\sim 8-10\,M_\odot$
for solar metallicity
(e.g., \citealt{Doherty2017PASA...34...56D}).
\citet{Limongi2024ApJS..270...29L} studied the 
evolution and final fate of solar metallicity MS stars 
with masses $\sim 7-15\,M_\odot$. They 
suggested that stars with masses $\sim 7.5-9.2\,M_\odot$ will develop degenerate ONe cores and evolve into the thermally pulsing super-AGB stage, in which stars with masses $\sim 8.5-9.2\,M_\odot$ 
may potentially explode as EC-SNe.

\begin{figure}
\begin{center}
\epsfig{file=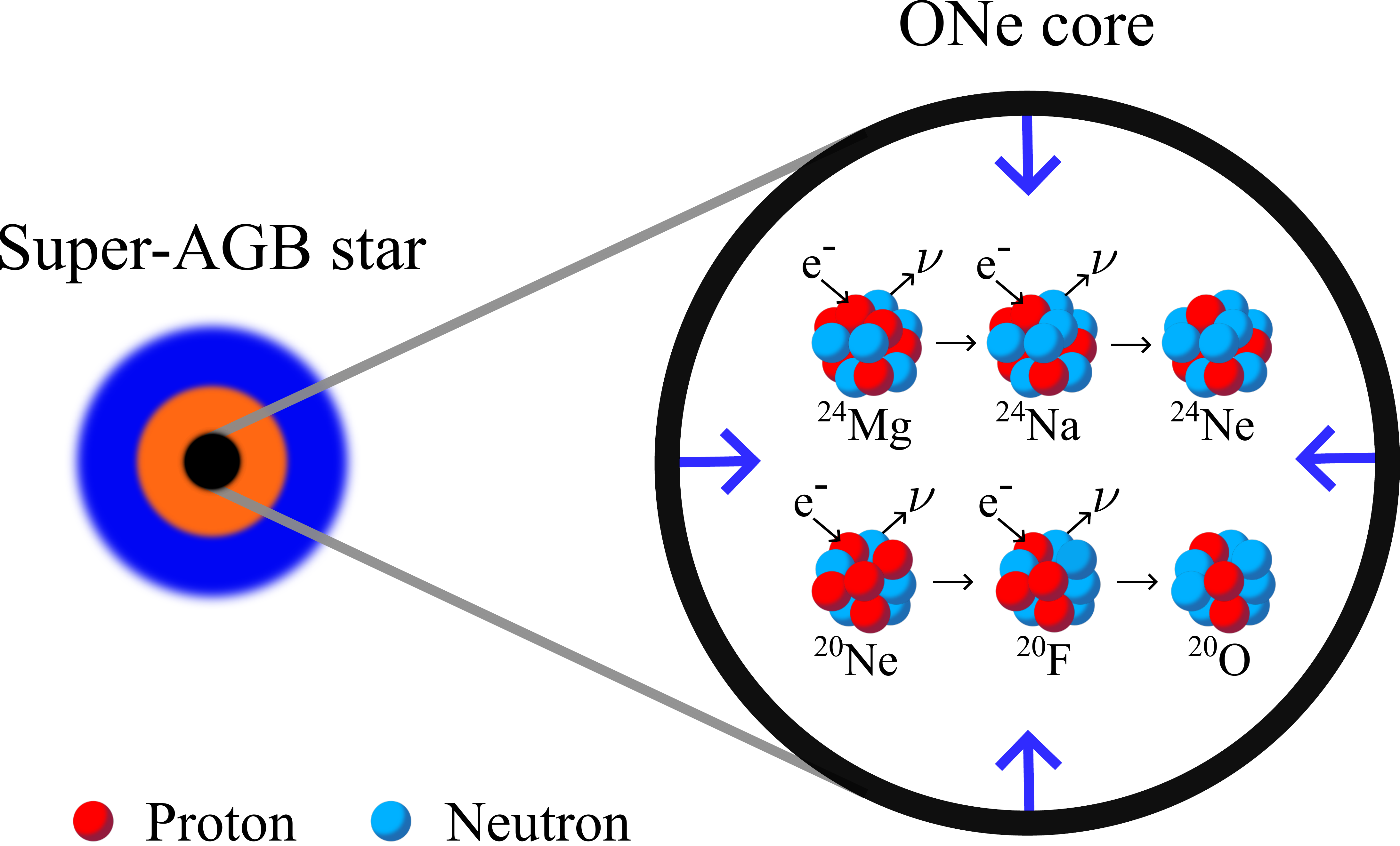,angle=0,width=7.5cm}
\caption{A super-AGB star and its degenerate ONe core that is supported by electron pressure. 
When the ONe core has mass close to $M_{\rm Ch}$, electron-capture reactions on $^{24}\rm Mg$ and $^{20}\rm Ne$ happen.
ln this process, Na and F are also formed along with lots of neutrinos.}
  \end{center}
  \label{fig:channel}
\end{figure}

\citet{WangNikki2024ApJ...969..163W} recently investigated the evolution of stars with  masses
$\sim 8.80-9.45\,M_\odot$ towards core collapse by electron-capture reactions. 
They found that the convective zone of the pre-collapse core can supply the required stochastic angular momentum fluctuations to set a jet-driven EC-SN explosion based on the jittering jets explosion mechanism (see \citealt{Soker2010MNRAS.401.2793S}).
Owing to uncertainties in the third dredge-up and the mass-loss rates in AGB stars, 
\citet{Poelarends2008ApJ...675..614P} roughly provided an upper limit to 
the contribution of super-AGB stars to all SNe ($\sim20\%$).
\citet{Parmar2026JHEAp..5000470P} recently argued that even a modest amount of asymmetric dark matter raises the central density of WD progenitors and lowers the mass threshold for EC-SNe,
allowing the formation of NSs with masses below the observed minimum.

Metallicity may affect the progenitor mass range for EC-SNe by 
influencing super-AGB stellar winds and the third dredge-up,
although substantial uncertainties persist in modeling these processes.
\citet{2007PhDT.......212P} suggested that 
the degenerate cores in super-AGB stars can grow up to higher masses with the decrease of metallicity,
thereby broadening the initial mass range for EC-SN progenitors, as follows: 
(1) For solar metallicity, EC-SNe have a narrow initial mass range of $\sim9.0-9.2\,M_\odot$.
(2) At the lowest metallicity ($Z=10^{-5}$),
all super-AGB stars would end life as EC-SNe with a corresponding initial mass range of $\sim6.4-8.2\,M_\odot$,
leaving no ONe WDs.
\citet{2007PhDT.......212P} argued that EC-SNe may account for about $5\%$ of all type IIP SNe at solar metallicity,
increasing up to $\sim38\%$ at $Z=10^{-5}$.
In contrast, by employing stellar evolution models,
\citet{Doherty2015MNRAS.446.2599D}  found a much narrow progenitor mass range of $\sim8.2-8.4\,M_\odot$ across different metallicities,
indicating that EC-SNe represent only a small fraction ($\sim2-5\%$) of the overall 
type IIP SN population.
It is still highly uncertain about the exact mass range of super-AGB stars for producing EC-SNe,
for more studies see, 
e.g., \citet{Nomoto1984ApJ}, \citet{Doherty2015MNRAS.446.2599D}, \citet{Farmer2015ApJS..220...15P}, \citet{Jones2016}, \citet{Zha2019ApJ...886...22Z}, \citet{LeungShing-Chi2019PASA...36....6L} and \citet{Leung2020ApJ...889...34L}, etc.

\subsection{He stars}

Aside from super-AGB stars, single He stars can also produce EC-SNe (see \citealt{Chanlaridis2022A&A...668A.106C}).
Massive stars can lose their H envelopes and become stripped He stars through some processes, such as binary interactions, common envelope (CE) ejection and strong stellar winds, etc
(see, e.g., \citealt{1995ApJ...448..315W,2017MNRAS.470.3970Y,2019ApJ...878...49W,2022A&A...661A..60A}).
\citet{Nomoto1984ApJ} studied the evolution of He stars towards EC-SNe, 
suggesting that the initial mass range for EC-SNe is between 2.0 and $2.5\,M_\odot$.
In the classic picture, it ranges from $1.37$ to $1.43\,M_\odot$ for 
the mass of the metal core in He stars required to produce EC-SNe 
(see, e.g., \citealt{Nomoto1984ApJ, Takahashi2013ApJ...771...28T, Tauris2015MNRAS.451.2123T}).

However,
\citet{2006PhDT.......214W} showed that isolated He stars can form degenerate ONe cores, where residual central carbon may trigger explosive oxygen burning.
\citet{Antoniadis2020A&A} investigated the detailed evolution of He star models with initial masses of $1.8-2.5\,M_\odot$ using the stellar evolution code MESA.
They suggested that the developed degenerate ONe cores initiate explosive oxygen burning at central densities $<6\times10^9\rm\,g\,cm^{-3}$ due to residual carbon burning,
thereby resulting in ONe type Ia SNe.
To investigate the role of residual central carbon, \citet{Antoniadis2020A&A} artificially turned off carbon burning after the ONe core formed.
This leads to electron-captures on $^{20}$Ne and subsequent oxygen ignition at a central density of $\sim10^{10}\rm\,g\,cm^{-3}$, resulting in the formation of an EC-SN rather than a type Ia SN.
Building on these findings,
\citet{Chanlaridis2022A&A...668A.106C} further simulated the evolution of He stars with initial masses in the range of $0.8-3.5\,M_\odot$ and metallicities between 0.0001 and 0.02,
and explored the effects of initial composition, stellar wind efficiency, and mixing assumptions, etc.
Their results support that
ONe type Ia SNe may be produced by degenerate cores with $M_{\rm Ch}$ retaining more than $\sim0.005\,M_\odot$ of residual carbon, whereas those with less carbon undergo EC-SNe
(see also \citealt{Guo2023MNRAS.526..932G}).
These results imply a narrow parameter space for the formation of EC-SNe through single He stars.

\section{The binary channel}
\subsection{He star in binaries}

Stellar evolution calculations indicate that the single star channel for EC-SNe is highly sensitive to the efficiency of the dredge-up process and the mass-loss rate,
with a very narrow progenitor mass window of only $\sim0.1-0.5\,M_\odot$ in width
(see, e.g., \citealt{2007A&A...476..893S,Poelarends2008ApJ...675..614P,Doherty2015MNRAS.446.2599D}).
However,
it is widely accepted that more than $70\%$ of massive stars reside in binaries (see, e.g., \citealt{Sana2012Sci...337..444S,2012MNRAS.424.1925C,Duchene2013ARA&A..51..269D,ChenXuefei2024PrPNP.13404083C}).
During the evolution of such systems, at least one component typically loses most of its H-rich envelope,
resulting in the formation of He star+MS or NS+He star binaries
(see, e.g., \citealt{Podsiadlowski2004ApJ...612.1044P, Tauris2015MNRAS.451.2123T}).
In this case, the dredge-up process can be avoided.

Recently, intermediate-mass stripped He stars in binaries have been confirmed through UV photometry (see \citealt{2023Sci...382.1287D}). In a further work,
\citet{2023ApJ...959..125G} presented the spectroscopic analysis for ten of such stripped stars, obtaining the stellar properties, such as effective temperature, surface gravity, masses and surface elemental abundance, etc.
Moreover,
\citet{Kruckow2018MNRAS.481.1908K} suggested that the first SN in the formation of a double NS system has a higher probability of being an EC-SN rather than an Fe CC-SN. 
Therefore, in this section we present an overview of EC-SNe occurring in He star+MS binaries and NS+He star binaries.

\subsubsection{He star+MS systems}

According to the stability of the first mass-transfer process, 
there are basically two evolutionary pathways for the formation of He star+MS binaries (see Fig.\,3).
These binaries are 
considered to produce X-ray binaries after the He stars experience EC-SNe, eventually forming double compact objects (e.g., \citealt{Tauris2006csxs.book..623T,2018A&A...614A..99S}), as follows:
\begin{enumerate}
    \item 
The stable mass-transfer scenario (Scenario A).
The primordial binary consists of two MS stars with comparable masses.
The primordial primary will fill its Roche-lobe when it evolves into the giant stage, leading to a stable mass-transfer process.
After that, the system becomes a binary composed of a He star and a massive MS companion.
The He star subsequently could undergo an EC-SN explosion, resulting in the formation of a NS. 
The system will evolve into a high-mass X-ray binary (HMXB) when the primordial secondary evolves into the giant stage and fills its Roche-lobe.
During this stage,  a CE could be formed if the mass-transfer is dynamically unstable.
After the CE ejection, a NS+He star system will be formed.
The second formed He star could also experience an EC-SN explosion if its ONe core grows in mass close to $M_{\rm Ch}$, 
after which the binary turns to be a double NS system (for details see Sect. 5.2).
    \item 
The CE evolution scenario (Scenario B).
In this case, the initial binary has a large mass ratio between the two components.
The system may experience a CE phase due to unstable mass-transfer process 
after the primordial primary fills its Roche-lobe at the giant stage.
After the CE ejection, the system becomes a close binary consisting of a He star and a low-mass MS companion.
The He star could explode as an EC-SN, forming a NS. 
The primordial secondary continues to evolve, and will fill its Roche-lobe when it becomes a giant star.
During the stable mass-transfer stage, the binary displays as a low-mass X-ray binary (LMXB).
After this stage, the secondary forms a WD. 
\citet{2015ApJ...809...99S} studied the formation of LMXBs by combining binary population synthesis (BPS)  with detailed stellar evolutionary calculations,
and considered different kick velocities for EC-SNe and  core-collapse SNe.
They found that, in the CE evolution scenario, the progenitor masses of NSs lie in the range of $7.6–8.6\,M_\odot$,
and the primordial binaries have initial orbital periods between $0.2$ and $1000\,$d.
\end{enumerate}

\begin{figure}
\begin{center}
\epsfig{file=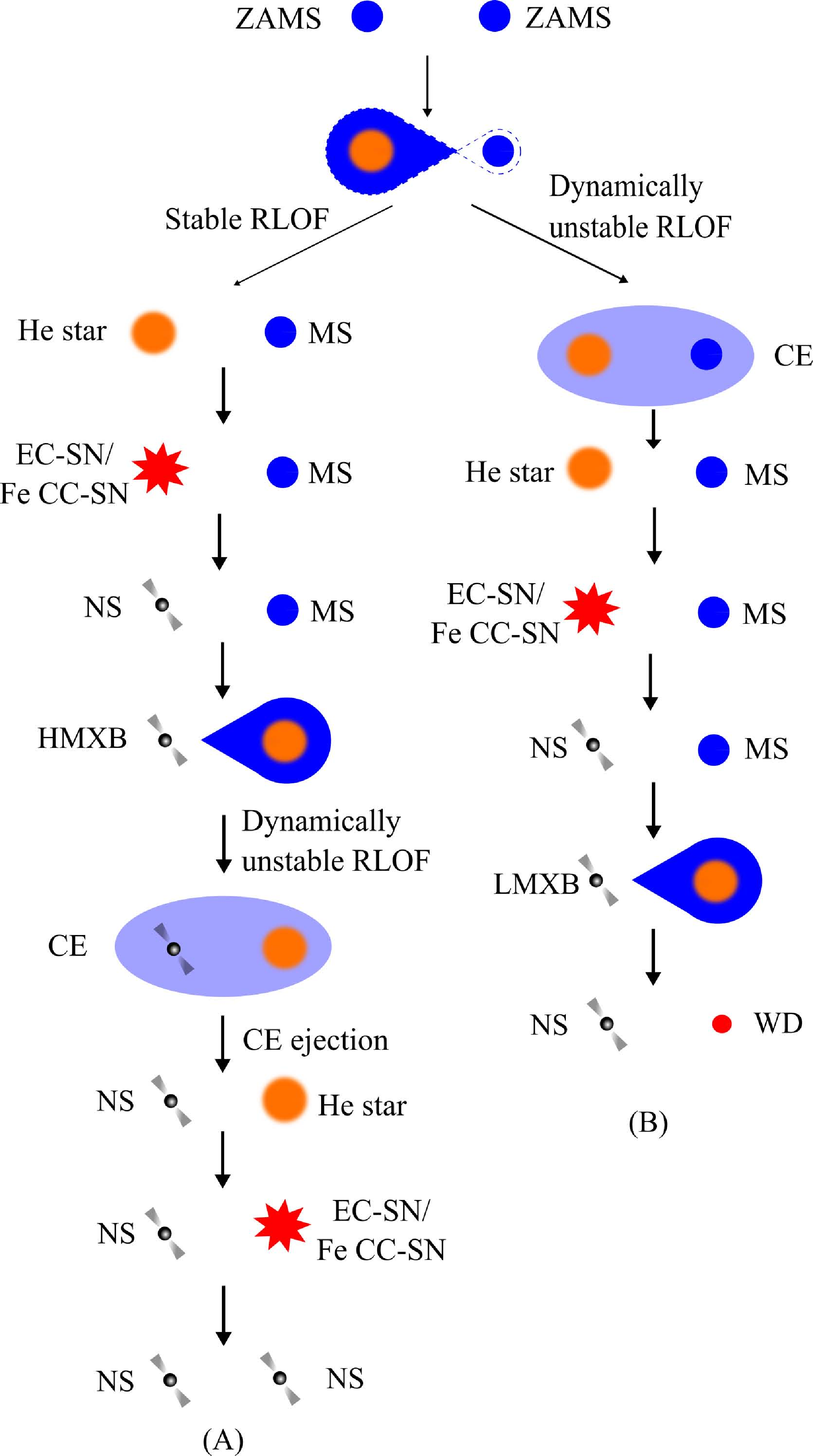,angle=0,width=7.5cm}
 \caption{Evolutionary scenarios for the formation of He star+MS systems, in which the He star could form an EC-SN (see also \citealt{Tauris2006csxs.book..623T, 2018A&A...614A..99S}). }
  \end{center}
  \label{fig:channel}
\end{figure}

\textbf{EC-SNe originating from Scenario A have been extensively studied.}
It is believed that the mass range for EC-SNe in binaries is significantly broader,
since the mass-loss through Roche-lobe overflow (RLOF) can help to avoid the second dredge-up that typically occurs in single star evolution.
The possibility of EC-SNe in binaries was first explored by \citet{Podsiadlowski2004ApJ...612.1044P}.
Based on the findings that He cores with masses in the range of $2.0-2.5\,M_\odot$ can undergo EC-SNe (see \citealt{Nomoto1984ApJ}), \citet{Podsiadlowski2004ApJ...612.1044P} proposed that the primary stars with  initial  masses between $8$ and $11\,M_\odot$ can evolve to EC-SNe in binaries.
\citet{2017ApJ...850..197P} conducted the first detailed modeling of EC-SNe in binaries using MESA,
suggesting that the initial primary masses for EC-SNe lie between $13.5$ and $17.6\,M_\odot$.
In their models,
CO cores with masses $\sim1.37-1.52\,M_\odot$ lead to EC-SNe owing to efficient cooling that suppresses Ne ignition below $1.52\,M_\odot$,
while more massive cores develop Ne burning and proceed to Fe CC-SNe. By adopting the criterion that EC-SNe can occur for metal cores with masses between $1.37$ and $1.43\,M_\odot$ (see \citealt{Tauris2015MNRAS.451.2123T}), however, \citet{2018A&A...614A..99S} argued that systems can produce EC-SNe within a relatively narrow primary mass range of $10.9-11.5\,M_\odot$.

\textbf{In Scenario A, the progenitors of EC-SNe can undergo either Case A or Case B evolution.}
(1)
Binary systems with an initial orbital period shorter than $\sim3\,$d undergo Case A mass-transfer, 
during which the primary is in the central H-burning stage (see \citealt{2017ApJ...850..197P}). However, 
\citet{2018A&A...614A..99S} argued that in all of their conservative Case A models the systems evolve into contact, rather than experiencing stable mass-transfer as reported in \citet{2017ApJ...850..197P}.
(2) 
Systems with initial orbital periods $\sim3-60\,$d undergo Case B mass-transfer, during which the primary is in the H-shell burning stage (see \citealt{2017ApJ...850..197P}).

\begin{figure}
\begin{center}
\epsfig{file=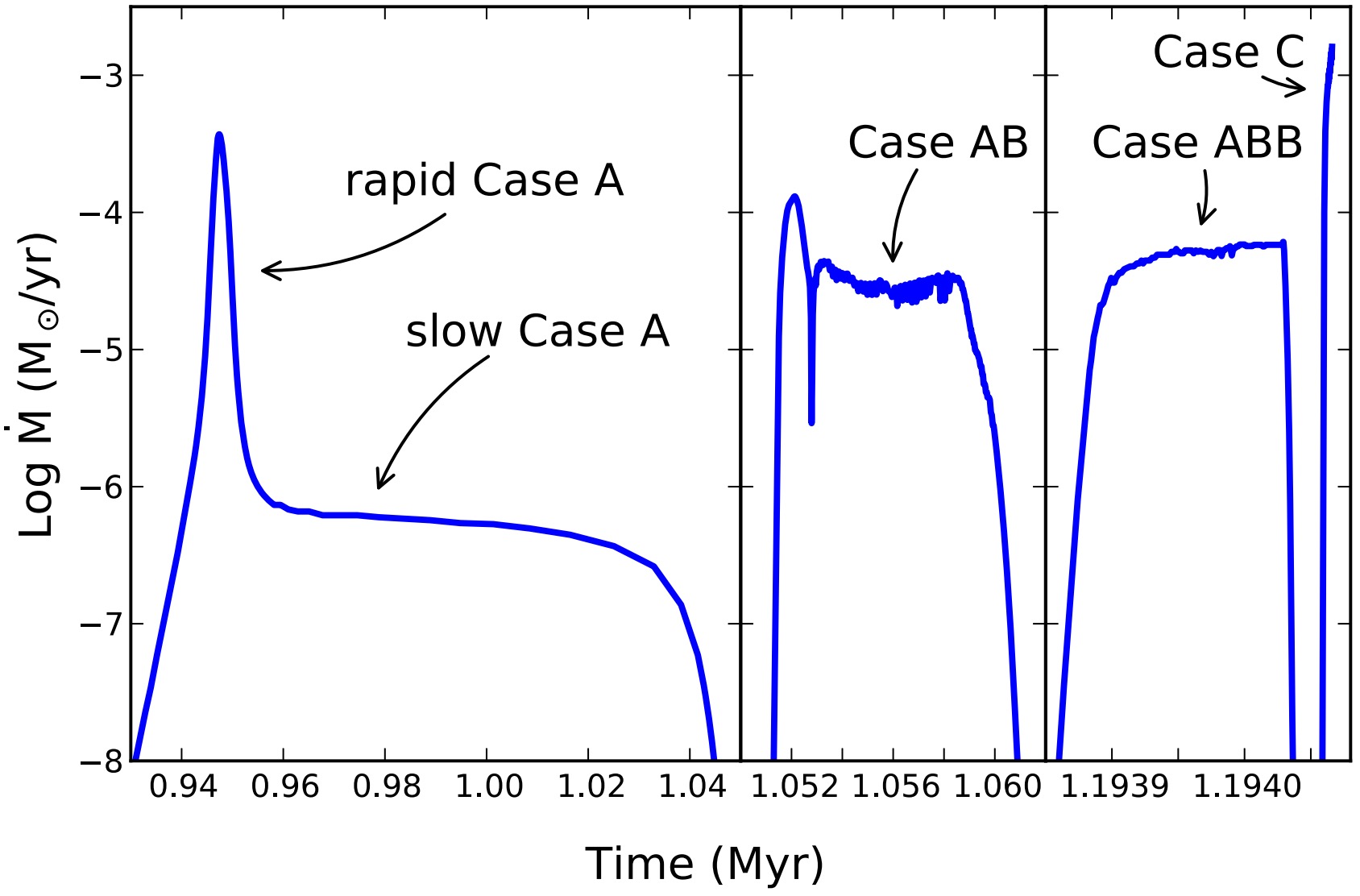,angle=0,width=9.5cm}
 \caption{The mass-transfer rate in the Case A system varies over time, exhibiting three clearly defined stages.
 The left panel illustrates the mass-transfer rate throughout the central H-burning stage. The middle panel displays the rates during H-shell burning stage, while the right panel presents the mass-transfer behavior during and following the carbon-burning stage.
 Source: From \citet{2017ApJ...850..197P}.}
  \end{center}
  \label{fig:caseamt}
\end{figure}

\textbf{A representative example for Case A evolution.} 
Fig. 4 shows the evolution of the mass-transfer rate
as a function of time in a system with an initial primary mass of $15.7\,M_\odot$, 
a secondary mass of $12.56\,M_\odot$ and an initial orbital period of $3\,$d (see \citealt{2017ApJ...850..197P}).
The first mass-transfer episode occurs on a thermal timescale, characterized by a high mass-transfer rate.
This stage is referred to as the rapid Case A mass-transfer
(e.g., \citealt{1994A&A...290..119P, 2001A&A...369..939W}),
during which the primary loses $\sim9\,M_\odot$.
After the mass ratio reversal, the mass-transfer rate decreases to $\sim10^{-6}\,M_\odot \rm yr^{-1}$, and the primary  evolves on a nuclear timescale.
Following the Case A mass-transfer, the primary is reduced to a total mass of $6.23\,M_\odot$ with a He core of $1.75\,M_\odot$.
Case AB mass-transfer is triggered by H-shell burning and proceeds on the thermal timescale of the star, with a high mass-transfer rate of $\sim3\times10^{-5}\,M_\odot \rm yr^{-1}$.
After the completion of Case AB mass-transfer, the primary has lost most of its H-envelope, and the binary becomes a He star+MS system.
The He star completes He core burning at $1.172\times10^7\,$yr.
Following this, core contraction triggers He-shell burning, causing the star to expand once more to red giant (RG) dimensions and initiate a new mass-transfer process, known as Case ABB.
By the onset of convective carbon burning in the core, the star has developed a CO core with a mass of $1.29\,M_\odot$.
After carbon burning,
the primary evolves into an ONe core with a mass of $1.19\,M_\odot$, covered by a $0.15\,M_\odot$ CO mantle (i.e., a $1.34\,M_\odot$ metal core).
At the end, the metal core has grown to $1.40\,M_\odot$ due to the ongoing He-shell burning, reaching the condition for triggering an EC-SN.
At this moment, 
the secondary acquires about $6\,M_\odot$, reaching a total mass of $18.58\,M_\odot$ and remaining on the MS stage.
If the primary undergoes an SN explosion and forms a NS, the system is likely to evolve into a Be/X-ray binary.

\textbf{A representative example for Case B evolution.}
For binaries with initial orbital periods of $\sim3-60\,$d, Case B mass-transfer is triggered when the primary fills its Roche lobe during different stages of the Hertzsprung gap,
leading to early, intermediate, or late Case B episodes (see \citealt{2017ApJ...850..197P}).
As the donor star expands rapidly on the thermal timescale during its crossing of the Hertzsprung gap, this results in a stable high mass-transfer process with rates of $\sim10^{-3}-10^{-2}\,M_\odot \rm yr^{-1}$.
After the Case B mass-transfer, the H-envelope of the primary is largely stripped and the system evolves into a MS+He star configuration.
The subsequent evolution resembles that of Case A systems, undergoing further mass-transfer during the He-shell burning stage (Case BB) and again during the C-shell burning stage (Case BBC).
With such successive stripping episodes, the primary is expected to form a massive ONe core,  eventually exploding as an EC-SN.

\subsubsection{NS+He star systems}
NS+He star binaries are typically formed after a CE phase 
triggered by unstable mass-transfer in HMXBs (see Scenario A in Fig. 3).
This kind of binaries are considered as the progenitors of double NSs that are   
important gravitational wave sources.
It seems that most double NSs are produced with relatively low kick velocities,
suggesting that EC-SNe play a significant role for the formation of double NSs
(see, e.g., \citealt{Tauris2017ApJ...846..170T, Shao2018ApJ...867..124S, Andrews2019ApJ...880L...8A, Guo2024MNRAS.530.4461G}).

\begin{figure}
\begin{center}
\epsfig{file=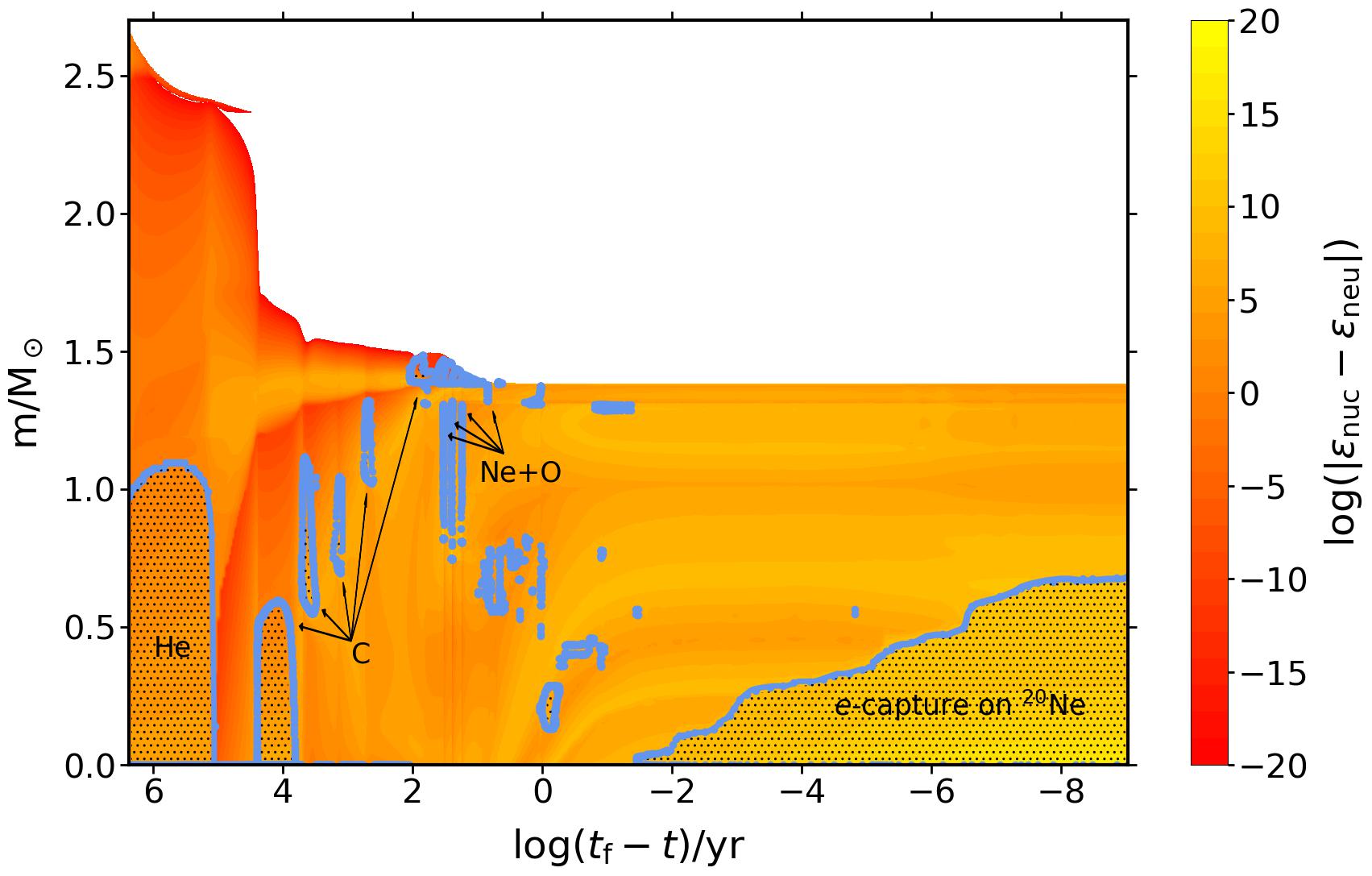,angle=0,width=11.5cm}
 \caption{Kippenhahn diagram of a $2.67\,M_\odot$ He star companion from He-ZAMS to electron-capture reactions,
including the evolution of interior structure and energy production (for the details of this evolution see \citealt{Guo2024MNRAS.530.4461G}).
The hatched regions denote convection caused by the He-, C-, and advanced burning phases.
The blue regions indicate the convection regions.
The intensity shown in the color-bar represents the nuclear energy-production rate.}
  \end{center}
  \label{fig:kip}
\end{figure}

\textbf{A representative example for binary evolution}. 
We present an example of the evolution of a NS+He star binary with an initial He star mass of $M_{\rm He}^{\rm i} = 2.67\,M_\odot$, an initial NS mass of $M_{\rm NS}^{\rm i} = 1.35\,M_\odot$ and an initial orbital period of $P_{\rm orb}^{\rm i}=1.0$\,d, in which the He star companion ultimately undergoes an EC-SN (see \citealt{Guo2024MNRAS.530.4461G}).
Fig. 5 shows the Kippenhahn diagram of the He star companion from the zero-age MS (ZAMS) to the onset of explosive oxygen burning, illustrating the evolution of its internal structure and energy production.
The total simulated stellar age from He-ZAMS to the onset of electron-capture is $2.37\,$Myr.
After central He exhaustion, a carbon–oxygen core develops, and subsequent He-shell burning causes the He star companion to expand and fill its Roche lobe, thereby triggering Case BB mass-transfer at $t\sim2.35\,$Myr.
The Case BB mass-transfer lasts for $\sim 0.03$\,Myr, during which the rates reach highly super-Eddington values, up to $\sim 10^3\dot M_{\rm Edd}$.
Previous studies suggested that NSs can accrete material at rates of 
$\sim2-3\,\dot M_{\rm Edd}$ during the Case BB mass-transfer (see, e.g., \citealt{2014MNRAS.437.1485L, Tauris2017ApJ...846..170T}).
\citet{Guo2024MNRAS.530.4461G} found that NSs may accrete even more mass if the residual H-rich envelope on the He star companion is taken into account.
Based on these results, we assume that NSs can accrete at a rate of $3\dot M_{\rm Edd}$ during the Case BB mass-transfer.
Under this assumption, the NS is expected to accrete about $2.7\times10^{-3}\,M_\odot$ of material,
which is sufficient to spin it up to become a millisecond pulsar (MSP) with a $\sim30$\,ms spin period.

\begin{figure}
\begin{center}
\epsfig{file=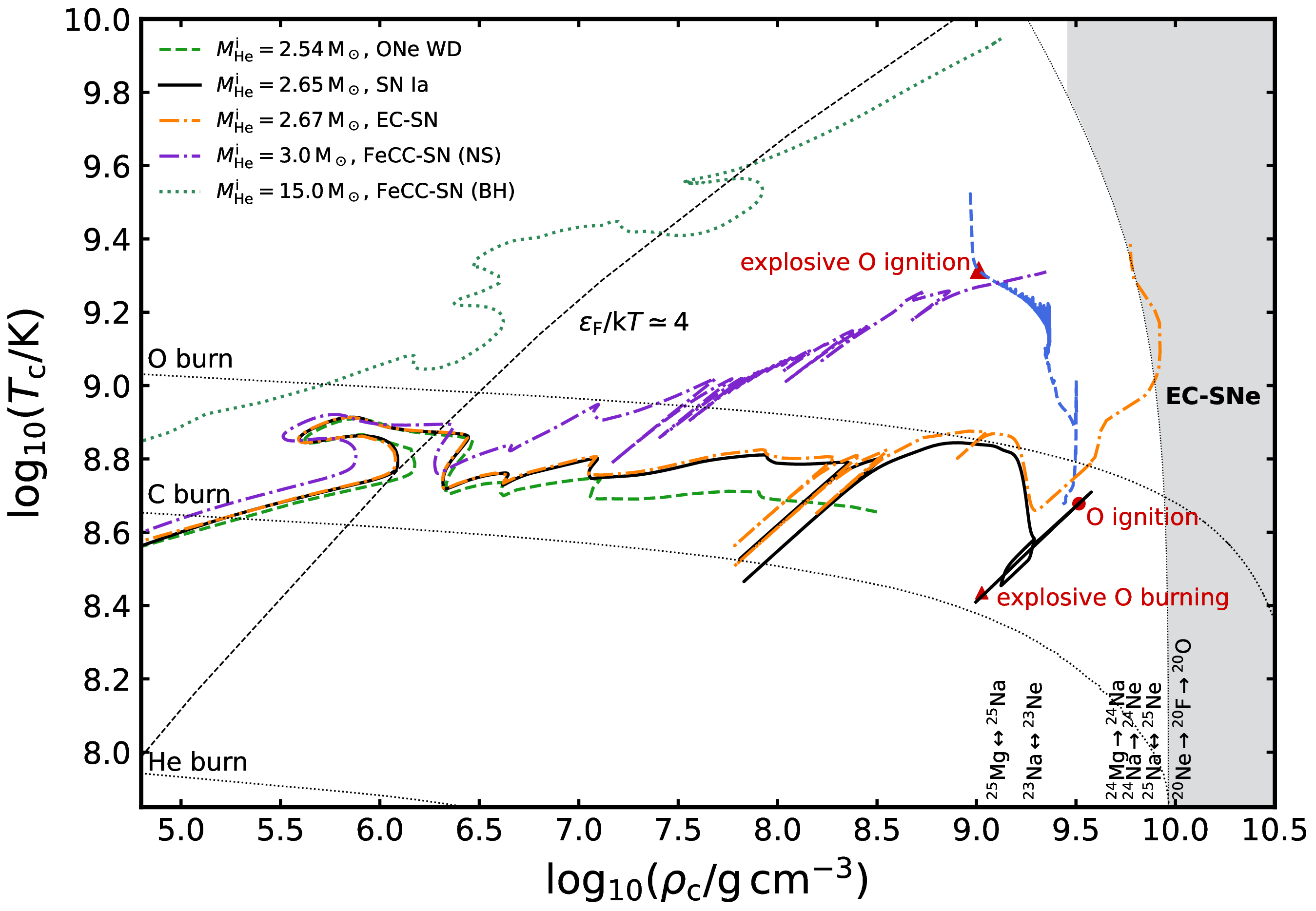,angle=0,width=11.5cm}
 \caption{Central temperature versus central mass density for NS+He star systems with different initial He star masses
(i.e., $M_{\rm He}^{\rm i}=2.54, 2.65, 2.67, 3.0$ and $15.0\,M_\odot$), in which the black dashed line indicates the boundary between degenerate and non-degenerate conditions (see also \citealt{Guo2023MNRAS.526..932G,Guo2024MNRAS.530.4461G}). In
all models, we assume an initial NS mass of $M_{\rm NS}^{\rm i} = 1.35\,M_\odot$ 
and an initial orbital period of $1.0$\,d.}
  \end{center}
\end{figure}

Fig. 6 represents the evolutionary track of the central density 
and temperature for NS+He star systems with different initial He star masses, in which 
different results are shown in this figure. 
For the case with $M_{\rm He}^{\rm i} = 2.67\,M_\odot$
(see the orange dash-dotted line), the He companion star will form an EC-SN
(see \citealt{Guo2024MNRAS.530.4461G}).
At the beginning, the He star companion develops an ONe core of $\sim1.385\,M_\odot$ following the carbon-burning stage, in which the ONe core is surrounded by a He-shell ($\sim0.1\,M_\odot$).
Subsequently, Ne is ignited off-center at a mass coordinate of $\sim0.8\,M_\odot$ due to a temperature inversion caused by neutrino emission in its core (i.e., a Ne-shell flash).
The central density and temperature increase after each Ne-shell flash is quenched
(see \citealt{2013ApJ...772..150J}).
After several such flashes, the central density rises to $\rm log_{10}(\rho_c/g\,cm^{-3})\sim9.1$, initiating Urca processes.
These reactions reduce the central temperature and accelerate the contraction of the metal core, thereby further increasing the central density.
As the central density gradually increases, electron-captures on $^{24}\rm Mg$ and $^{20}\rm Ne$ occur at $\rm log_{10}(\rho_c/g\,cm^{-3})\sim9.6$ and
$\sim9.9$, respectively.
Meanwhile, a thick Si-rich mantle is formed as a result of the Ne-shell flashes.
At this stage, the remaining He envelope has a final mass of only $\sim8.2\times10^{-4}\,M_\odot$,
and thus He lines are not expected to be visible in the post-explosion spectra (see \citealt{2012MNRAS.422...70H}).

\begin{figure}
\begin{center}
\epsfig{file=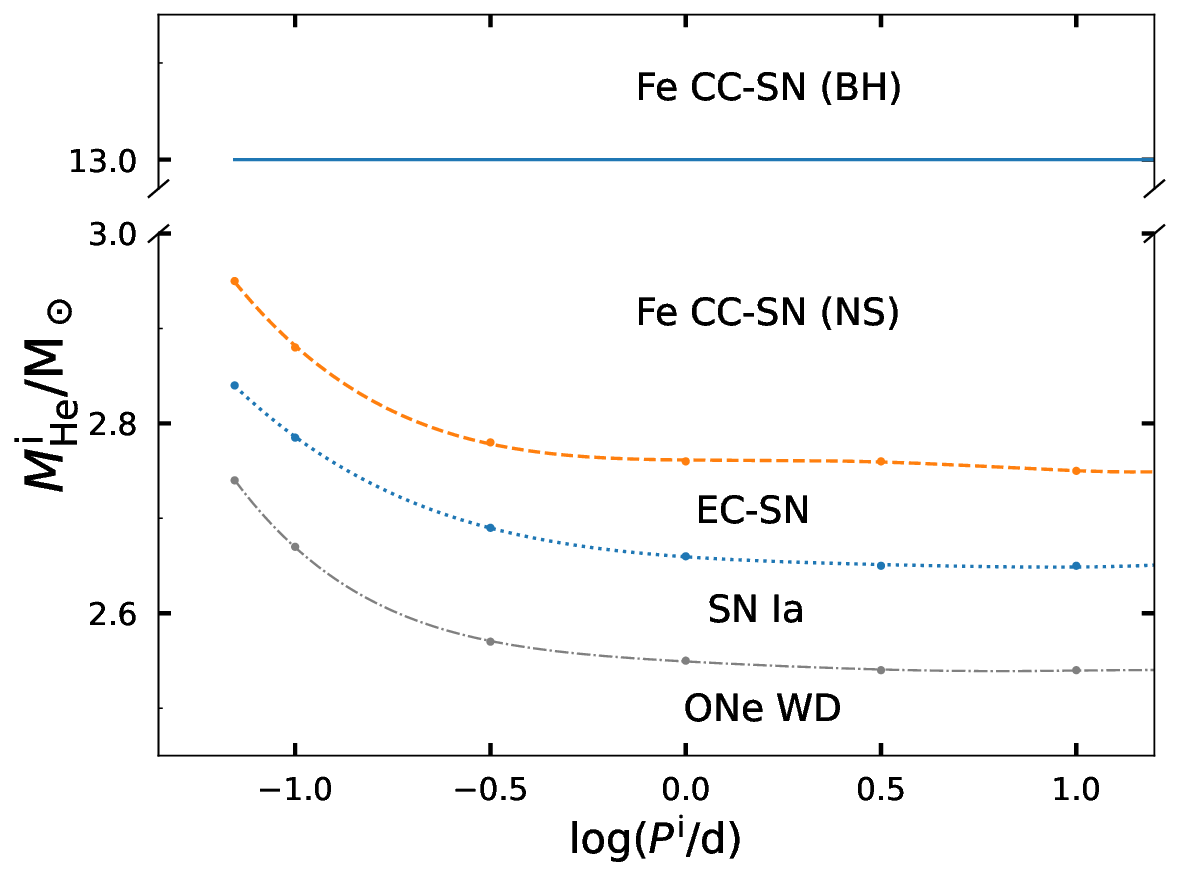,angle=0,width=11.5cm}
 \caption{Initial parameter spaces in the $\rm log\,$$P^{\rm i}_{\rm orb}-M^{\rm i}_{\rm He}$ plane for NS+He star binaries, in which we set the initial NS mass as $M_{\rm NS}^{\rm i} = 1.35\,M_\odot$.
 Systems with initial orbital periods lager than $10$\,d yield similar outcomes with single He stars. 
 The different results are indicated in this figure (see also \citealt{Guo2023MNRAS.526..932G,Guo2024MNRAS.530.4461G}). }
  \end{center}
\end{figure}

\citet{Tauris2015MNRAS.451.2123T} investigated the evolution and final outcomes of NS+He star systems with initial orbital periods ranging from $0.06$ to $\rm 2\,d$  and initial He star masses between $2.5$ and $3.5\,M_\odot$.
They provided the parameter space leading to the formation of CO WDs, ONe WDs, EC-SNe, and Fe CC-SNe.
For solar metallicity,
they found that the  mass range for He stars required to produce EC-SNe lies between $2.60$ and $2.95\,M_\odot$, depending on the initial orbital period.
\citet{Guo2024MNRAS.530.4461G} recently conducted a systematic study of EC-SNe in NS+He star systems. 
Fig. 7 shows the initial parameter spaces of NS+He star binaries, 
in which the  He star companions will produce different kinds of objects with the decrease of their initial masses,
i.e., Fe CC-SNe (BHs), Fe CC-SNe (NSs), EC-SNe, SNe Ia and ONe WDs.
The left boundaries of the parameter spaces are determined by the condition that RLOF occurs while the He star companions are still on the He-ZAMS stage. If the initial orbital period is $\gtrsim10$\,d, the He star companions tend to follow similar evolutionary paths and final outcomes.
\citet{Guo2024MNRAS.530.4461G} also found that both the initial orbital period and the minimum initial orbital period for EC-SNe increase with metallicity.
In Sect. 5.2, we will discuss the formation of double NSs through NS+He star binaries, 
in which the He star companion experiences an EC-SN.

\subsection{Accretion-induced collapse of WDs}
Aside from the He star in binaries, 
accretion induced collapse (AIC) of oxygen-neon (ONe) WDs in binaries can also form a kind of EC-SNe, eventually producing NS binaries or isolated NSs. 
ONe WDs are suggested to collapse into NSs through electron-capture reactions 
when they grow in mass close to ${M}_{\rm Ch}$,
called the AIC process  (see, e.g., \citealt{Nomoto+1979wdvd.coll...56N,Taam+1986ApJ...305..235T,Nomoto1991ApJ...367L..19N}). 
There exits a lot of indirect evidence to support the AIC process that
may help to explain some peculiar NS systems in observations owing to small kicks (see, e.g., \citealt{Canal1990ARA&A..28..183C,Li1998ApJ...500..935L,Ivanova2008MNRAS.386..553I,Chen2011MNRAS.410.1441C,Boyles2011ApJ...742...51B,Tauris2013A&A...558A..39T,Ablimit2015ApJ...800...98A,Ablimit2019ApJ...881...72A}).
Although the AIC process was proposed as a final fate of ONe WDs,
there is no direct detection for such events to date.
Up to now, two classes of formation ways have been proposed to produce AIC events, i.e.,
the single-degenerate scenario and the double-degenerate scenario (for a recent review see \citealt{Wang+2020RAA....20..135W}).
Recent progress on the two progenitor scenarios is reviewed, containing the evolutionary ways from the primordial binaries to AIC events, the initial parameter contours for AIC events, 
and the pre/post-AIC systems in the observations (see \citealt{Wang+2020RAA....20..135W}).

\subsubsection{The single-degenerate scenario}

In the single-degenerate scenario,
an ONe WD increases its mass by accreting H-/He-rich matter from a
non-degenerate companion, in which the mass donor could be a 
MS star (the  MS donor channel), 
a RG star (the  RG donor channel), or  a He star (the  He star donor channel).
An AIC event may be formed
once the ONe WD increases its mass approach ${M}_{\rm Ch}$
(see, e.g.,  \citealt{Nomoto1991ApJ...367L..19N,Ivanova+2004ApJ...601.1058I,Tauris2013A&A...558A..39T,Brooks2017ApJ...843..151B,Wang2018MNRAS.481..439W,LiuD2018MNRAS.477..384L,Ruiter2019MNRAS.484..698R,ZhangZ2024ApJ...975..186Z}). 

For the MS donor channel, the pre-AIC systems  
are likely to be identified  as supersoft X-ray sources, 
recurrent novae and Ne novae when the ONe WD accretes H-rich matter.\footnote{U Sco is a recurrent nova that has a $1.55\pm0.24\,M_{\odot}$ WD and a $0.88\pm0.17\,M_{\odot}$ MS donor
with an orbit period $\sim0.163$\,d (e.g., \citealt{Hachisu2000ApJ...528L..97H,Thoroughgood2001MNRAS.327.1323T}).
\citet{Mason2013A&A...556C...2M} pointed out that the mass-accretor in  U Sco is possibly a CO WD, resulting in a type Ia SN finally.
However, \citet{Schaefer2025ApJ...991..110S} recently argued that 
the WD in U Sco  is losing large masses each eruption cycle based on its orbital period changes, indicating that U Sco can never form a type Ia SN.}
After the AIC process, the MS donor may fill its Roche-lobe again, 
and transfer H-rich material onto the newborn pulsars.
The post-AIC systems with MS donors could be identified as the LMXBs, resulting in the formation of the fully recycled MSPs with He WD companions.
\citet{Ablimit2015ApJ...800...98A} suggested that the post-AIC systems with MS donors could 
reproduce the properties of LMXBs with the strong magnetics like 4U 1822$-$37.
It has been argued that the AIC of magnetic WDs with MS donors could form  two types of 
eclipsing MSPs, i.e., black widows and redbacks
(see \citealt{Ablimit2019ApJ...881...72A}).

For the  RG donor channel, the pre-AIC systems could be identified 
as symbiotic novae in the observations, such as T CrB,
RS Oph, J0757 and  V745 Sco, etc (e.g., \citealt{Belczynski1998MNRAS.296...77B,Brandi2009A&A...497..815B,Tang2012ApJ...751...99T,Mikolajewska2017ApJ...847...99M,Orlando2017MNRAS.464.5003O}).
It is worth noting that the type of WDs in these symbiotic novae is still not well determined.
In a subsequent evolution, the post-AIC systems with RG donors could be identified as 
LMXBs, producing the more mildly recycled MSPs 
with wide orbits finally ($>30$\,d; see \citealt{Wang2022MNRAS.510.6011W}).

For the  He star donor channel, the pre-AIC systems could be identified as 
supersoft X-ray sources and  He novae  during the mass-transfer process.
The post-AIC systems with He star donors could be identified as 
intermediate-mass X-ray binaries, eventually forming intermediate-mass binary pulsars (IMBPs) with short orbits. 
\citet{LiuD2018MNRAS.477..384L} investigated the He star donor channel for the formation of IMBPs. 
They found that this channel may explain most of the observed IMBPs with short orbits ($<$3\,d), 
especially for the formation of PSR J1802$-$2124.
Recent studies indicate that CO WD+He star systems can also 
form ONe WD+He star systems when off-center carbon burning happens 
on the CO WD,
eventually producing NS systems through the AIC process 
(see \citealt{Brooks2016ApJ...821...28B,Wang+2017MNRAS.472.1593W}).

\subsubsection{The double-degenerate scenario}

In the double-degenerate scenario, AIC events originate from the merging of double WDs 
with a total mass larger than ${M}_{\rm Ch}$. 
The merging of double WDs was caused 
owing to orbit shrinking resulting from gravitational wave radiation, forming
isolated NSs finally (e.g., \citealt{Nomoto1985ApJ...297..531N,Saio1985A&A...150L..21S,Ruiter2019MNRAS.484..698R,Liu2020MNRAS.494.3422L}).
This scenario mainly includes four channels, i.e., the double CO WD channel, 
the double ONe WD channel, the ONe WD+CO WD channel, and the ONe WD+He WD channel (see \citealt{Wang+2020RAA....20..135W}).

The double CO WD channel has been thought to be one of the two classic ways for producing type Ia SNe 
(e.g., \citealt{Webbink1984ApJ...277..355W, Iben1984ApJS...54..335I, Han1998MNRAS.296.1019H}). However,
off-center carbon burning may happen on the massive CO WD 
owing to a relatively high mass-transfer rate during the merging process, 
resulting in the formation of ONe WDs via an inwardly propagating carbon flame
 (e.g., \citealt{Nomoto1985ApJ...297..531N,Saio1985A&A...150L..21S,Timmes1994ApJ...420..348T}).  
It has been suggested that the AIC process may be avoided when the merging process is violent, in which
the mass-ratio of the two CO WDs is lager than 0.8 (e.g., \citealt{Pakmor2010Natur.463...61P,Pakmor2011A&A...528A.117P,Sato2016ApJ...821...67S}).
According to a thick-disc assumption (i.e., the slow merger model),
\citet{Wu2019MNRAS.483..263W} recently suggested that 
the final outcomes of double CO WDs  
strongly depend on the mass-transfer rates during the merging.
Some possible progenitor candidates for this channel have been discovered, 
such as KPD 1930+2752 (e.g., \citealt{Maxted2000MNRAS.317L..41M,Geier2007A&A...464..299G,Liu2018MNRAS.473.5352L}), 
Lan 11 (e.g., \citealt{LuoChangqing2025SCPMA..6869511L}),
HD 265435 (e.g., \citealt{Pelisoli2021NatAs...5.1052P,Qi2023RAA....23a5008Q}), 
WDJ181058.67+311940.94 (e.g., \citealt{Munday2025NatAs...9..872M}), 
and Henize 2-428 (e.g., \citealt{Santander2015Natur.519...63S,Wu2019RAA....19...57W}), etc.
It is noting that 
the outcomes of double CO WDs are still highly uncertain (see, e.g., 
\citealt{Yoon2007MNRAS.380..933Y,Kromer2013ApJ...778L..18K,Bulla2016MNRAS.455.1060B,YuY2019ApJ...870L..23Y,Li2023A&A...669A..82L,Liu2023RAA....23h2001L,Ruiter2025A&ARv..33....1R,Shen2025ApJ...982....6S}).

For the double ONe WD channel, it is generally believed that
the merger remnant will collapse into a NS via the electron-capture reactions (e.g., \citealt{Miyaji+1980,Schwab2015MNRAS.453.1910S}).
Recent studies indicate that the outcomes of the merger remnant
may arise from the competition between the electron-capture reactions and the thermalnuclear runaway of O/Ne 
(e.g., \citealt{Wu2018RAA....18...36W,Marquardt2015A&A...580A.118M,Jones2016,Jones2019A&A}).  
\citet{WuChengyuan2023MNRAS.525.6295W} recently 
simulated the remnant evolution of double ONe WD
merger, in which the remnant ignites off-center O/Ne burning soon after the merger. 
They found that the final fates of the merger
remnants are sensitive to the process of convective boundary mixing; 
if the mixing process cannot prohibit the O/Ne flame from reaching the
center, the merger remnants would undergo Fe CC-SNe to form NSs,
otherwise their final fates may be ONeFe WDs via EC-SNe. 

For the ONe WD+CO WD channel, 
\citet{Dan2014MNRAS.438...14D} investigated the evolution of ONe+CO WD merger remnants, 
suggesting that the merger remnants might evolve into a kind of unusual type Ia SNe  
or a kind of faint type Ic SNe.
\citet{Kashyap2018ApJ...869..140K} argued that the merger remnants
might form relatively faint and rapidly fading type Ia SNe   
via a failed detonation, fainter even than the faintest type Iax events.
Recent simulations suggested that the final fates of such merger remnants are related to their initial masses;  
more massive remnants ($>1.95\,M_{\odot}$) would become EC-SNe and form
CO-rich circumstellar material (CSM) resulting from the stellar wind during their giant stage, in which off-center Ne burning occurs soon after the merger, whereas  less massive remnants ($<1.90\,M_{\odot}$) would experience a  carbon-shell burning stage to increase the core mass before off-center Ne burning (see \citealt{Wu2023ApJ...944L..54W}).
\citet{Wu2023ApJ...944L..54W} found that the merging of ONe+CO WD systems can reproduce
the main observational properties of a hot star (i.e., J005311) in the constellation Cassiopeia. 
In a subsequent work, \citet{WuChengyuan2024ApJ...967L..45W} found that the merging of ONe WD+CO WD 
can explain the observed properties of a peculiar type Icn SN, i.e., SN 2019jc.

For the ONe WD+He WD channel,
\citet{Brooks2017ApJ...850..127B} carried out the evolutionary simulations of 
ONe WD+He WD merger remnants, and they suggested that the 
ONe WD cores would experience EC-SNe after growing to near 
${M}_{\rm Ch}$.   
\citet{Liu2023MNRAS.521.6053L} recently evolved a series of ONe
WD+He WD systems for the formation of AIC events, eventually producing a type of peculiar NS binaries 
like 4U 1626–67 that consists of a newly formed NS and an ultra-light companion star. It has been suggested that an ultra-compact X-ray binary 
(i.e., XTE\,J1751$-$305) 
may originate  from the evolution of an ONe WD+He WD system  (see \citealt{Liu2023MNRAS.521.6053L}).

\section{Candidates of EC-SNe}


EC-SNe are expected to explain some peculiar low luminosity SNe or events. 
Depending on the mass loss during the late evolutionary stage of their progenitors,
EC-SNe may exhibit diverse objects in observations, 
including normal to faint  type IIP SNe
(e.g., \citealt{Botticella2009MNRAS.398.1041B, Hosseinzadeh2018ApJ...861...63H,Hiramatsu2021NatAs...5..903H}),
type IIn SNe (see \citealt{Moriya2014AA...569A..57M}),
type IIn-P SNe (type IIn SNe with an early plateau phase as seen in type IIP SNe; see \citealt{Smith2013MNRAS.434..102S}),
type Ib/c SNe (see \citealt{Tauris2015MNRAS.451.2123T}),
intermediate-luminosity red transients
(e.g., \citealt{2021A&A...654A.157C, Rose2025ApJ...980L..14R})
and rapidly evolving faint/luminous transients (see, e.g., \citealt{Moriya2016MNRAS.461.2155M,Tolstov2019ApJ...881...35T}),
etc.

Some possible candidates for EC-SNe have been proposed, e.g., 
SN 1997D (a low luminosity type IIP SN; see, e.g., \citealt{Turatto1998ApJ...498L.129T,Chugai2000A&A...354..557C,Benetti2001MNRAS.322..361B}),
SN 2005cs (a low luminosity type IIP SN; see, e.g., \citealt{Pastorello2006MNRAS.370.1752P,Pastorello2009MNRAS.394.2266P}),
SN 2008S (a faint type IIn SN; see, e.g., \citealt{Botticella2009MNRAS.398.1041B,Thompson2009ApJ...705.1364T,Pumo2009ApJ...705L.138P}),
SN 2008ha 
(the faintest SN 2002cx-like object; see, e.g., \citealt{Pumo2009ApJ...705L.138P,Moriya2016MNRAS.461.2155M}),
KSN 2015K (a rapidly evolving luminous SN; see \citealt{Tolstov2019ApJ...881...35T}),
SN 2015bf (a fast declining type II SN; see \citealt{LinHan2021MNRAS.505.4890L}),
SN 2016bkv (a low luminosity type IIP SN; see \citealt{Hosseinzadeh2018ApJ...861...63H}),
SN 2018zd (a typical type IIP SN; see, e.g., \citealt{Zhang2020MNRAS.498...84Z,Hiramatsu2021NatAs...5..903H,Callis2021arXiv210912943C})
and
AT 2019abn (an intermediate-luminosity red transient; see \citealt{Rose2025ApJ...980L..14R}), etc.

Among the candidates of EC-SNe, 
SN 2018zd is a promising one that has strong observational evidence for EC-SN origin, including the
progenitor identification, the explosion energy,  the light curve, the CSM, the chemical composition 
and the nucleosynthesis, etc
(see \citealt{Hiramatsu2021NatAs...5..903H,Sato2024ApJ...970..163S}; see also \citealt{Zhang2020MNRAS.498...84Z}).
\citet{Hiramatsu2021NatAs...5..903H} inferred that the progenitor of SN 2018zd is a super-AGB star.  
However, \citet{Kozyreva2021MNRAS.503..797K} argued that the photometric and spectroscopic evolution of SN 2018zd closely resembles that of a typical type IIP SN, suggesting instead a low- to intermediate-mass Fe CC-SN
(see also \citealt{Callis2021arXiv210912943C}). 

Aside from SN 2018zd, some EC-SN candidates may also have super-AGB stars as the progenitors, as follows:
(1) \citet{Moriya2014AA...569A..57M} suggested that SN 2008S could originate from an EC-SN if the H-rich envelope in an super-AGB star is small. 
(2) \citet{Tolstov2019ApJ...881...35T} argued that KSN 2015K could be formed through
an optically thick CSM around an super-AGB star.
(3) Owing to a low nickel mass as well as dusty CSM,  
\citet{LinHan2021MNRAS.505.4890L} suggested that 
SN 2015bf could originate from an EC-SNe through a super-AGB star.
(4) According to the James Webb Space Telescope spectrum, 
\citet{Rose2025ApJ...980L..14R} recently suggested that 
AT 2019abn has a super-AGB progenitor and explodes as an EC-SN.
In addition, \citet{Chugai2000A&A...354..557C} speculated that the progenitor of SN 1997D have initial MS masses of 
$\sim 8-12\,M_\odot$, and 
the progenitor of SN 2005cs has been suggested to be a red supergiant  with an initial mass of $\sim 7-13\,M_\odot$ (see \citealt{Pastorello2009MNRAS.394.2266P} and references therein).
It is worth noting  that SN 2008ha may be related to 
stripped-envelope EC-SNe 
that are predicted as rapidly evolving faint transients with relatively small velocities (see \citealt{Moriya2016MNRAS.461.2155M}).

Additionally, it has been
suggested that Crab Nebula (the remnant of the historical SN 1054) may
originate from the collapse of a super-AGB star as
an EC-SN,
mainly due to the low inferred kinetic energy of the filaments 
and the observed peculiar compositions (e.g., abundant He and less abundant O) in its remnant
(see \citealt{Miyaji+1980,Nomoto1982Natur.299..803N,Stockinger2020MNRAS.496.2039S,Horvath2022Ap&SS.367...81H,Sato2024ApJ...970..163S,Limongi2024ApJS..270...29L}). 
\citet{Smith2013MNRAS.434..102S} pointed out that
type IIn-P SNe caused by electron-capture explosions
can reproduce several observed properties of the Crab Nebula.
Furthermore,
\citet{Moriya2014AA...569A..57M} found that 
the explosion of an EC-SN within an ordinary super-AGB wind 
can explain the observed light curve features of SN 1054.
However, both the progenitor and the explosion mechanism for SN 1054 are still widely under debate (see \citealt{Omand2025MNRAS.536..408O}).

\section{Impacts of EC-SNe on some fields}

EC-SNe have important impacts on some astrophysical fields, for example, the properties for NSs originated 
from EC-SNe, the formation of double NSs and the chemical enrichment of galaxies, etc.

\subsection{Properties of NSs}

The birth mass function of NSs contains
a lot of information about SN explosion, binary evolution, and 
the equation of state for the matter under extreme conditions, etc. 
It is a long-standing issue for the determination of the NS mass distribution in astrophysics.
\citet{Thorsett1999ApJ...512..288T} proposed that the observed NS masses are in a narrow range
with Gaussian mass distribution ($M=1.35\pm0.04\,M_\odot$).
Meanwhile, it has been suggested that there are two distinct populations for 
the mass distribution of NSs in observations, i.e., 
a low-mass population ($\sim 1.25\,M_\odot$) and a high-mass population ($\sim 1.35\,M_\odot$);
the low-mass population is the result of EC-SNe with small kicks, whereas the high-mass population originates in Fe CC-SNe with large kicks (see, e.g., \citealt{Timmes1996ApJ...457..834T, Schwab2010ApJ...719..722S}).
By collecting a sample of 90 NSs with well-determined mass estimates,
\citet{YouZhiQiang2025NatAs...9..552Y} recently argued that 
the birth masses of NSs can be reproduced by a distribution that peaks at $\sim1.27\,M_\odot$ before declining as a steep power law. 

The birth mass of the NS is determined by the mass cut during the explosion and the NS equation of state, both of which are still poorly constrained.
By considering a probable reduction of 10\% gravitational mass
during the collapse into NSs (see \citealt{Hudepohl2010PhRvL.104y1101H}), 
the NS masses produced in EC-SNe are expected in the range of $\sim 1.24-1.29\,M_\odot$ (see \citealt{Kruckow2018MNRAS.481.1908K}).
In addition,
PSR J0737$-$3039 is the first double pulsars discovered, consisting of Pulsar A (with spin periods of 22.7\,ms) and Pulsar B (with spin periods of 2.77\,s) (see \citealt{Burgay2003Natur.426..531B,Lyne2004Sci...303.1153L}).
According to the observed orbital parameters combined with the likely evolutionary history of PSR J0737$-$3039,  
\citet{Podsiadlowski2005MNRAS.361.1243P} inferred that the Pulsar B with masses $\sim1.249\pm0.001\,M_\odot$ may originate from an EC-SN explosion.

Owing to low kick velocities,
EC-SNe can be used to explain the retention of NSs in globular clusters and 
a group of HMXBs with wide orbits and low eccentricities 
(e.g., \citealt{Pfahl2002ApJ...573..283P,Podsiadlowski2004ApJ...612.1044P,vanden2010NewAR..54..140V}). 
Meanwhile, the AIC process in WD binaries may help to solve the large observed discrepancy 
between MSPs and LMXBs (the progenitors of MSPs) in the Galaxy
(e.g., \citealt{Bailyn1990ApJ...353..159B,Tauris2013A&A...558A..39T}).

\subsection{Double NSs}

Double NSs are important potential gravitational wave sources,
which can be produced in NS+He star systems after the He star companions experience EC-SNe or Fe CC-SNe   
(e.g., \citealt{Dewi2002MNRAS.331.1027D,Ivanova2003ApJ...592..475I,Dewi2003MNRAS.344..629D,Andrews2015ApJ...801...32A, Tauris2015MNRAS.451.2123T,Tauris2017ApJ...846..170T,Beniamini2016MNRAS.456.4089B,Kruckow2018MNRAS.481.1908K,Jiang2021ApJ...920L..36J,Guo2024MNRAS.530.4461G}).
Up to now, about 24 double NSs have been discovered after the first detection of double NSs five decades ago
(PSR B1913$+$16; \citealt{Hulse1975ApJ...195L..51H}).\footnote{ATNF Pulsar Catalogue
(version 2.6.5, August 2025; Manchester et al. 2005), available at \href{http://www.atnf.csiro.au/research/pulsar/psrcat}{http://www.atnf.csiro.au/research/pulsar/psrcat}.}
In the observations, 
the component masses of double NSs are mainly in the range of $1.1-1.5\,M_\odot$, and
the orbital periods have a wide range from $0.1$ to $45.0$\,d 
with the orbital eccentricities  from $0.06$ to $0.80$
(see \citealt{Manchester2005AJ....129.1993M}).
The observed properties of double NSs provide important clues for exploring their progenitors,
such as NS masses, spin periods, orbital periods and eccentricities, etc
(for previous studies see \citealt{Tauris2017ApJ...846..170T,Shao2018ApJ...867..124S,Andrews2019ApJ...880L...8A}).

\begin{figure}
\begin{center}
\epsfig{file=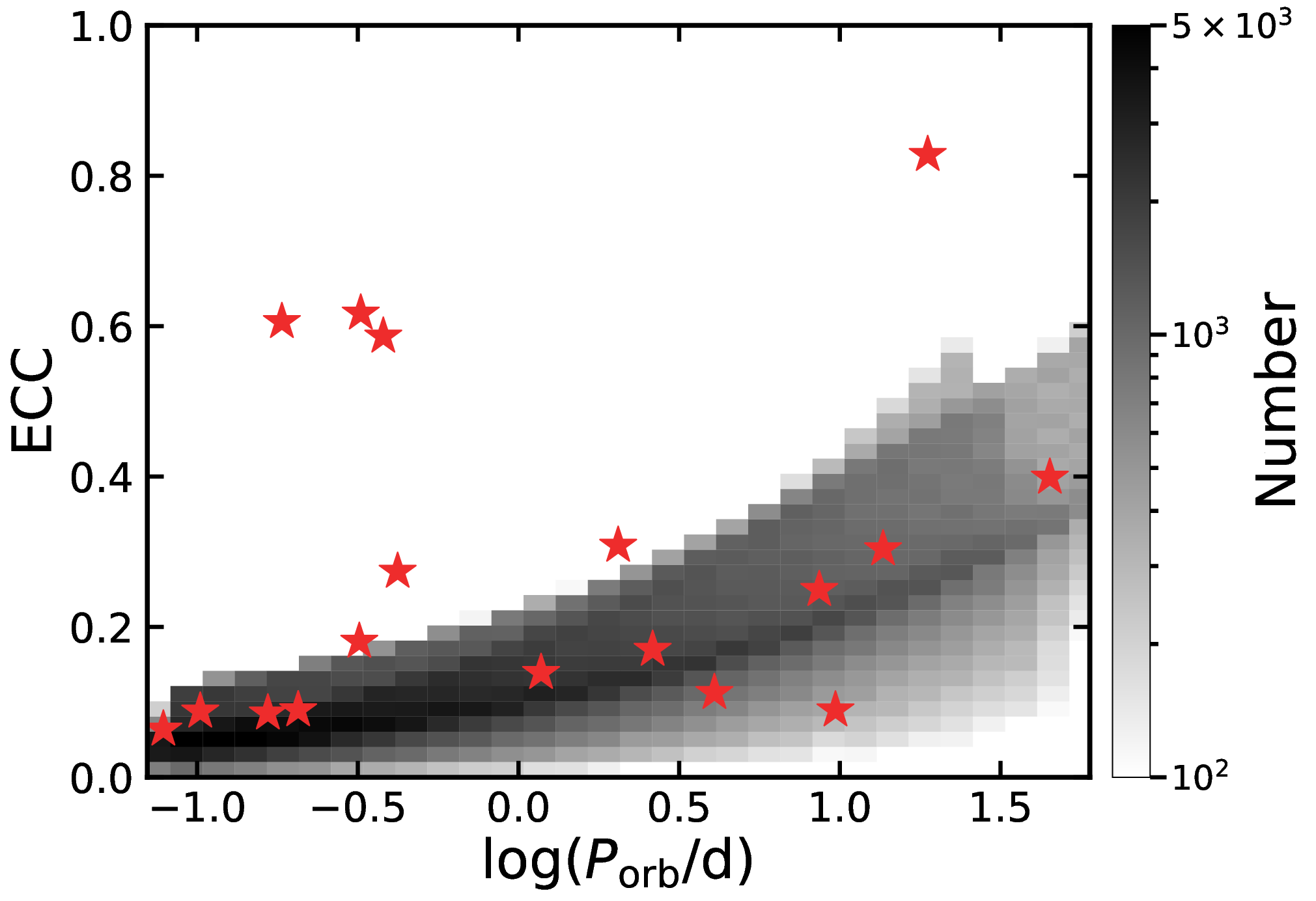,angle=0,width=11.5cm}
 \caption{
 Distribution of the simulated double NSs in the $P_{\rm orb}-e$ plane based on the NS+He star systems, in which the He star explodes as an EC-SN.
 We consider $500$ pre-SN orbital periods logarithmically spaced from $0.07$ to $50\,$d,
 a NS kick velocity of $30\,$km/s, and an ejected mass of $0.08\,M_\odot$.
 Each pre-SN system is evolved through $1000$ times with randomly oriented kicks (see also \citealt{Guo2024MNRAS.530.4461G}).
 The red stars show the observed Galactic double NSs, with data taken from the ATNF Pulsar Catalogue (version 2.6.5).}
  \end{center}
  \label{fig:kick}
\end{figure}

Fig. 8 shows the distribution of the simulated double NSs in the $P_{\rm orb}-e$ diagram based on the NS+He star systems, in which the He star explodes as an EC-SN.
It can be seen that NS kicks of $\lesssim30\,$km/s can reproduce the properties of most double NSs,
especially binaries with relatively low eccentricities or wide systems with relatively high eccentricities.
Previous studies suggested that most observed double NSs may be formed with small NS kicks ($\lesssim50\rm\,km^{-1}$),
implying that EC-SNe provide an important pathway to produce double NSs (e.g., \citealt{Tauris2017ApJ...846..170T, Shao2018ApJ...867..124S, Guo2024MNRAS.530.4461G}).

PSR $\rm J0453+1559$ has been proposed as a peculiar candidate for double NSs, 
characterized by an orbital period of $4.07\,$d and an eccentricity of $0.1125$
(see \citealt{2015ApJ...812..143M}).
The measured masses of the recycled pulsar and its companion are
$1.559(5)\,M_\odot$ and $1.174(4)\,M_\odot$, respectively.
While the nature of the companion remains uncertain, its relatively low mass compared to typical second-formed NSs in known double NSs makes it less likely to be a NS.
However, its eccentric orbit is unusually high for a typical NS+WD system.
To reconcile these features,
\citet{Tauris2019ApJ...886L..20T} proposed that the system may originate from a thermonuclear 
EC-SN with a resulting natal kick in the range of
$\sim70-100\rm\,km\,s^{-1}$.
In this scenario, ONe cores may avoid collapsing into NSs and instead leave bound ONeFe WD remnants
(see \citealt{Jones2016, Jones2019A&A,Kirsebom2019PhRvL.123z2701K}).
In order to verify this scenario,
more numerical simulations and observations for thermonuclear EC-SNe are required.

\subsection{Chemical products}

The nucleosynthesis outcome of EC-SNe is markedly different from that of typical Fe CC-SNe.
It has been suggested that EC-SNe can be as 
the possible mechanism for producing elements of $r$-process nucleosynthesis, 
contributing to the chemical evolution of galaxies 
(e.g., \citealt{Ning2007ApJ...667L.159N}).
Previous studies indicate that EC-SNe can explain the origin of
neutron-rich isotopes like $^{48}\rm Ca$, $^{50}\rm Ti$, $^{54}\rm Cr$ and $^{60}\rm Fe$
(see \citealt{Wanajo2013ApJ...767L..26W,Wanajo2013ApJ...774L...6W}).
In a further study,  \citet{Wanajo2018ApJ...852...40W} pointed out that
EC-SNe could be one of the major origins for light trans-Fe elements from Zn to Zr in the Galaxy,
which may help to  explain the observed abundances in some metal-poor stars
(e.g., \citealt{2018ApJ...855...63H}).

\citet{WangTianshu2024ApJ...969...74W} 
recently argued that the explosion energies in 3D models could be 2$-$10 times higher than that in  1D models, and correspondingly large differences in the $^{56}\rm Ni$ yields. 
They also suggested that the ejecta are more neutron-rich based on 3D models, 
leading to significant weak $r$-process and $^{48}\rm Ca$ yields.
Compared to typical Fe CC-SNe, EC-SNe contribute negligibly to the overall Fe-group elements
but may play a non-negligible role in shaping the Galactic inventory of specific neutron-rich isotopes.
The exact yields, however, remain uncertain because they depend sensitively on the electron fraction ($Y_{\rm e}$) distribution,
the neutrino transport treatment, and the details of the explosion mechanism
(e.g., \citealt{Wanajo2018ApJ...852...40W}).
For more studies on the contribution of EC-SNe to the galactic chemical evolution, see, e.g., \citet{Wanajo2011ApJ...726L..15W}, \citet{Takahashi2013ApJ...771...28T}, 
\citet{Nomoto2017hsn..book..483N}, \citet{Jones2019ApJ...882..170J} and \citet{LeungShing-Chi2019PASA...36....6L}, etc.

\section{Compared with ultra-stripped SNe}

Ultra-stripped SNe are defined as exploding stars that retain extremely 
small envelope masses ($\lesssim0.2\,M_\odot$; e.g., \citealt{Tauris2015MNRAS.451.2123T}).
As a kind of core-collapse SNe, ultra-stripped SNe are deficient in H (type Ib, IIb SNe) and possibly He (type Ic SNe) in their spectra (see \citealt{Filippenko1997}).
The progenitors of ultra-stripped SNe are likely stripped of their envelope via a combination of stellar winds and interactions with their companions (see \citealt{Haynie2023ApJ...956...98H}).
These events are characterized by extremely small ejecta masses,
rapidly evolving light curves, and very low natal kicks
($\lesssim50\,$km/s),
with strong observational candidates including SN 2005ek, iPTF14gqr, SN 2019dge, SN 2021agco  and SN 2023zaw, etc
(e.g., \citealt{2013ApJ...774...58D, 2013ApJ...778L..23T, 2018Sci...362..201D, 2020ApJ...900...46Y, YanShengyu2023ApJ...959L..32Y,Das2024ApJ...969L..11D,Moore2025ApJ...980L..44M}).
They may collapse through either EC-SNe or Fe CC-SNe,
although the majority of ultra-stripped SNe are expected to be Fe CC-SNe because EC-SNe require a narrow range of initial core masses for He stars (see \citealt{Tauris2015MNRAS.451.2123T}).
Ultra-stripped SN explosions are astrophysically significant
because they are believed to be a major  channel for the formation of double NSs.
Their low mass ejection and weak natal kicks greatly increase the likelihood
that binaries remain bound after the second formed SN,
helping to explain the observed population of double  NSs 
and the systems that eventually merge as gravitational wave sources
(e.g., \citealt{Tauris2017ApJ...846..170T, Kruckow2018MNRAS.481.1908K, Jiang2021ApJ...920L..36J,2025arXiv250800186C}).

Although ultra-stripped SNe and EC-SNe share some 
similar observable properties (e.g., low ejecta masses and small kicks),
their progenitor systems and collapse mechanisms differ in systematic ways.
Ultra-stripped SNe typically arise in very compact interacting binaries with orbital periods $\lesssim2\,$d,
usually consisting of a compact object (a NS, a BH or a massive WD) and a He star,
where extreme stripping through mass transfer removes nearly the entire envelope prior to collapse
(e.g., \citealt{Tauris2015MNRAS.451.2123T, 2020Sci...367..577V}).
The progenitors of ultra-stripped SNe therefore span a relatively narrow evolutionary path that requires a close binary and efficient stripping.
EC-SNe, by contrast,
originate from progenitors whose degenerate ONe cores grow toward the critical mass for collapse ($\sim1.43\,M_\odot$).
These include super-AGB stars, signle He stars,
stripped He stars in binaries, and AIC of WDs in interacting binaries.
The progenitor parameter space for EC-SNe is therefore considerably broader
than that of ultra-stripped systems,
encompassing both single and binary evolution pathways
that produce an ONe core.

In terms of explosion mechanisms,
ultra-stripped SNe are predominantly Fe core–collapse events,
in which photo disintegration of heavy nuclei and rapid electron-captures
reduce the pressure support of the Fe core and accelerate its collapse,
followed by neutrino-driven shock revival
(e.g., \citealt{2012ARNPS..62..407J, 2020MNRAS.491.2715B}).
EC-SNe, on the other hand, undergo collapse when rapid electron-captures on $^{20}$Ne and $^{24}$Mg
sharply reduce the electron pressure in a degenerate ONe core,
leading to a prompt collapse characterized by lower explosion energies, lower ejecta masses and a distinct nucleosynthetic signature, etc.
\citet{Sato2024ApJ...970..163S} recently
presented the synthetic light curves of EC-SNe and low-mass Fe CC-SNe, 
suggesting that EC-SNe have brighter, bluer, and shorter plateaus than those of Fe CC-SNe.
For more studies on the difference between EC-SNe and low-mass Fe CC-SNe, 
see, e.g., \citet{Moriya2014AA...569A..57M}, \citet{Tauris2015MNRAS.451.2123T}, 
\citet{Takahashi2019ApJ...871..153T}, \citet{Jones2019ApJ...882..170J}, \citet{Kozyreva2021MNRAS.503..797K} and \citet{Kozyreva2022MNRAS.514.4173K}, etc. 

\section{Summary and perspective}

EC-SNe are expected to originate from electron-capture reactions 
happening in ONe cores with masses close to $M_{\rm Ch}$, 
relating to the formation of isolated NSs and NS systems.
At present, the most studied pathways for EC-SNe 
are the single star channel (involving super-AGB stars and He stars)  
and the binary channel (involving He star in binaries and AIC in WD binaries). 
In this review, 
we outline recent progress the two primary progenitor channels in a systematic way, involving both single and binary stars. We also discuss the EC-SN candidates,
the broader astrophysical impacts of EC-SNe
on some research fields and
the difference between ultra-stripped SNe and EC-SNe in this article.

The study of EC-SNe is entering a critical stage where both observational and theoretical
advances can provide decisive tests of current models.
Upcoming large-scale surveys,
such as those conducted by the Vera C. Rubin Observatory (LSST),
are expected to dramatically increase the discovery rate of faint and rapidly evolving transients, potentially uncovering new EC-SN candidates in nearby galaxies
(see \citealt{2019ApJ...873..111I}).
Although promising candidates such as SN 2018zd have been proposed,
the reliable identification of EC-SNe among their diverse manifestations and 
the robust distinction from low-mass Fe CC-SNe remain central challenges.
Follow-up observations with facilities like the James Webb Space Telescope (JWST)
and next-generation 30 meter class optical/infrared telescopes
will enable high-resolution spectroscopy of low-luminosity events,
constraining ejecta composition, progenitor environments, and nucleosynthetic signatures such as the Ni/Fe ratio
(e.g., \citealt{2015RAA....15.1945S, 2018sf2a.conf....3N, 2023PASP..135f8001G}).
Multi-wavelength monitoring, combined with state-of-the-art radiative transfer modeling (see, e.g., \citealt{2018MNRAS.475..277J}),
will help establish robust criteria for distinguishing EC-SNe from low-mass Fe CC-SNe.
In addition,
next-generation gravitational wave detectors
(e.g., Einstein Telescope and Cosmic Explorer) may provide complementary constraints by
probing the formation of low-mass NSs potentially produced through EC-SNe
(e.g., \citealt{2019BAAS...51g..35R, 2020JCAP...03..050M}).

On the theoretical side,
key challenges remain in accurately modeling the evolution and collapse of super-AGB stars.
In particular,
the progenitor initial mass range leading to EC-SNe is still poorly constrained,
being highly sensitive to the treatment of stellar wind mass-loss rates and the efficiency of the third dredge-up.
Furthermore,
the propagation of the ONe flame and the physical conditions required to trigger collapse into a NS
depend on weak nuclear reaction rates, convective processes,
the formulation of turbulent flame speed, and the treatment of the laminar deflagration phase
(e.g., \citealt{Jones2016, Leung2020ApJ...889...34L}).
To resolve these issues,
it is essential to develop self-consistent multidimensional simulations of the collapse and explosion,
including improved turbulent flame physics and accurate modeling of convective boundary mixing,
which are critical for determining whether an ONe core undergoes electron-capture induced collapse.
Together, advances in theoretical modeling and progenitor characterization
will help establish a comprehensive understanding of EC-SNe
and their role in producing NSs and faint transients in the Universe.

\begin{acknowledgements}
The authors would like to thank the referee for valuable comments that helped improve this review.
The authors also acknowledge
Adam Burrows, Shuai Zha, Jujia Zhang, Luhan Li, Chengyuan Wu,
Yongzhi Cai and Zhuling Deng  
for their helpful discussions and suggestions. 
This study is supported by the National Natural Science Foundation of China 
(Nos 12225304, 12288102, 12090040/12090043, 12403035 and 12273105), the National Key R\&D Program of China (No. 2021YFA1600404), 
the Yunnan Revitalization Talent Support Program (Yunling Scholar Project, 
Innovation Team and Young Talent Project), the Yunnan Science and Technology Program (Nos. 202501AS070005, 202201BC070003, 202401AV070006 and 202201AW070011), and the International Center of Supernovae, Yunnan Key Laboratory (No. 202302AN360001).
\end{acknowledgements}

\bibliography{wang}
\bibliographystyle{raa}

\end{document}